\documentclass[twocolumn,aps,reprint,showpacs,preprintnumbers,amsmath,amssymb,groupedaddress]{revtex4-1}
\usepackage{graphicx}
\usepackage{dcolumn}
\usepackage{bm}
\usepackage{epstopdf}

\usepackage{amsmath}
\usepackage{subfigure}

\usepackage{soul}
\usepackage[normalem]{ulem}
\usepackage{color}

\begin{document}


\title{Resonant enhancement in nanostructured thermoelectric performance via electronic thermal conductivity engineering}

\author{Urvesh Patil}
\affiliation{%
Department of Electrical Engineering,
Indian Institute of Technology Bombay, Powai, Mumbai-400076, India\\
}%

\author{Bhaskaran Muralidharan}%
\email{bm@ee.iitb.ac.in}
\affiliation{%
Department of Electrical Engineering, Indian Institute of Technology Bombay, Powai, Mumbai-400076, India\\
}%



\date{\today}

\begin{abstract}
The use of an asymmetric broadening in the transport distribution, a characteristic of resonant structures, is proposed as a route to engineer a decrease in electronic thermal conductivity thereby enhancing the electronic figure of merit in nanostructured thermoelectrics. Using toy models, we first demonstrate that a decrease in thermal conductivity resulting from such an asymmetric broadening may indeed lead to an electronic figure of merit well in excess of $1000$ in an idealized situation and in excess of $10$ in a realistic situation. We then substantiate with realistic resonant structures designed using graphene nano-ribbons by employing a tight binding framework with edge correction that match density functional theory calculations under the local density approximation. The calculated figure of merit exceeding $10$ in such realistic structures further reinforces the concept and sets a promising direction to use nano-ribbon structures to engineer a favorable decrease in the electronic thermal conductivity.
\end{abstract}
\maketitle
\section{Introduction}
Low-dimensional systems \cite{dresselhaus1,dresselhaus2,Chen_1,sny08,sofo,heremans,nak} and nanostructures \cite{Chen_2, akshay,Chen_1} are envisioned as promising directions en route to the enhancement of the thermoelectric figure of merit. The thermoelectric figure of merit, $zT$, is defined as $$zT=\frac{S^2 \sigma T}{\kappa_{el}+\kappa_{ph}},$$ where $S$ is the Seebeck coefficient, $\sigma$ is the electronic conductivity $\kappa_{el}$ is the electronic thermal conductivity and $\kappa_{ph}$ is the lattice thermal conductivity. The term $S^2 \sigma$, appearing in the numerator is referred to as the power factor, and it relates to the actual electrical power that can be drawn by the load \cite{Nemir}. While much of the work on $zT$ enhancement has focused on engineering a decrease in the lattice thermal conductivity \cite{Chen_1,Chen_2,Gunst2011,Mazzamuto2011,Sevincli2010,Feng2016,Xie2016}, electronic engineering that aims to enhance the electronic figure of merit $z_{el}T=\frac{S^2 \sigma T}{\kappa_{el}}$, is somewhat a less explored direction \cite{shakouri}. In this context, the traditional direction followed is that of increasing the power factor via electron filtering \cite{faleev,shakouri} in low-dimensional structures \cite{dresselhaus1,dresselhaus2,nak,shakouri}.\\
\begin{figure}[t]
	\subfigure[]{\includegraphics[width=2in, height=1.2in]{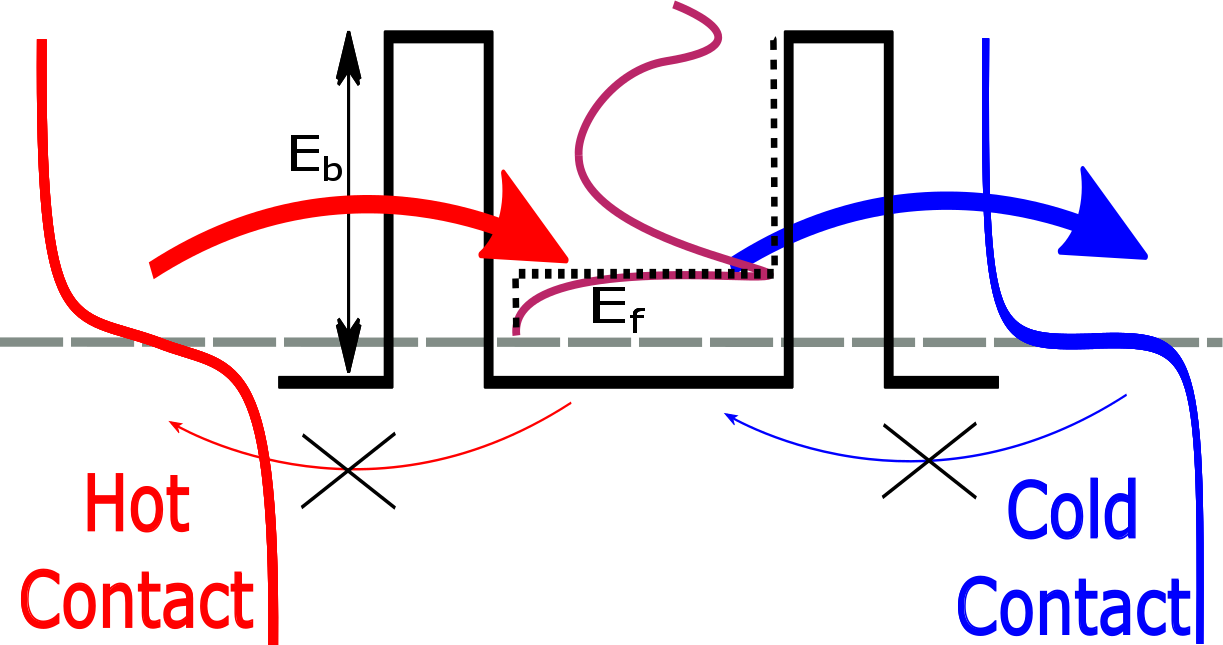}
	}
	
	\subfigure[]{\includegraphics[width=2in, height=1.2in]{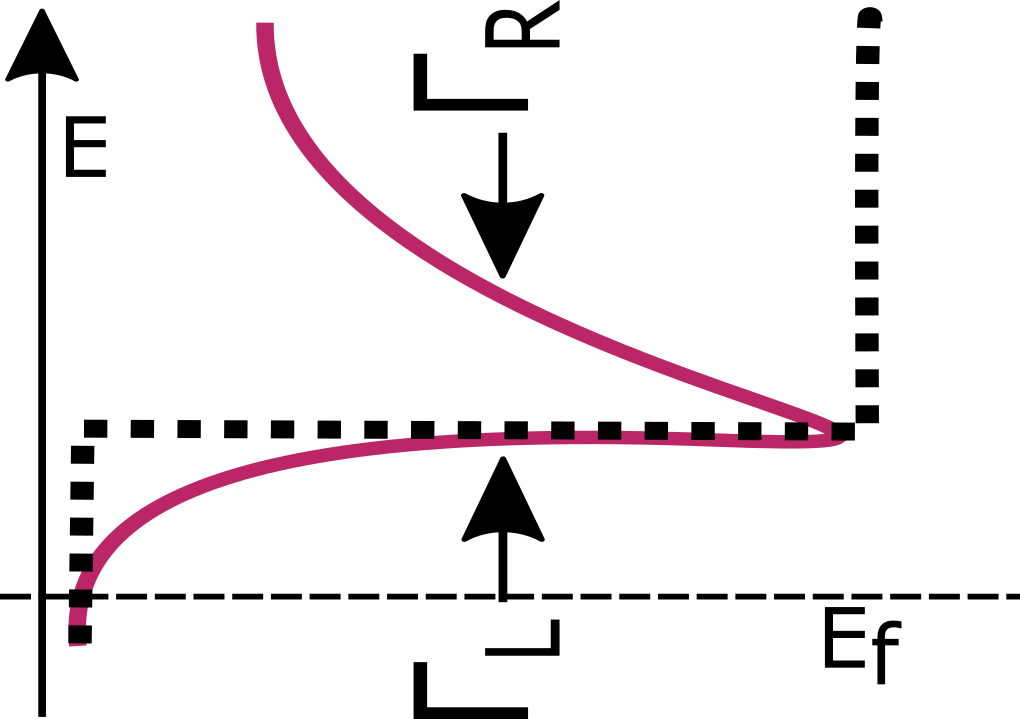}
	}
	\caption{Asymmetric broadening basics. (a) Schematic of a resonant tunnelling device depicting electron filtering effect due to suppressed flow of electrons from cold (right) to hot (left) contact as a result of sharp gradients in the asymmetry of the transmission function about the peak at the energy center $E_0$. (b) Typical asymmetric transmission obtained for a RTD device (solid). The dashed line depicts the transmission function of a typical two dimensional (2-D) structure that demonstrates an ideal electron filtering set up.}
	\label{fig:RTD}
\end{figure}
\begin{figure}[t]
 \subfigure[]{\includegraphics[width=1.7in, height=1.2in]{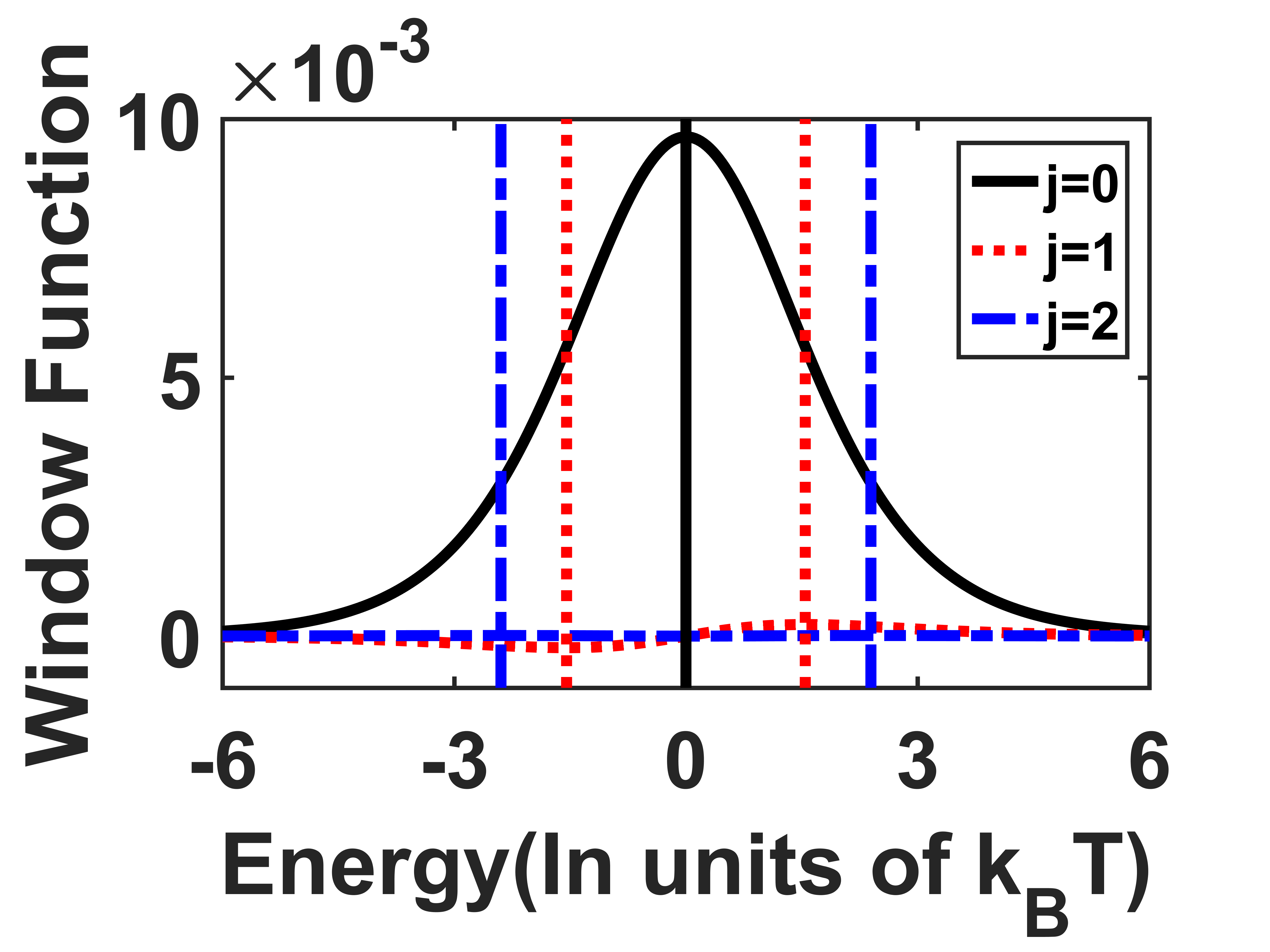}
 }\subfigure[]{\includegraphics[width=1.7in, height=1.2in]{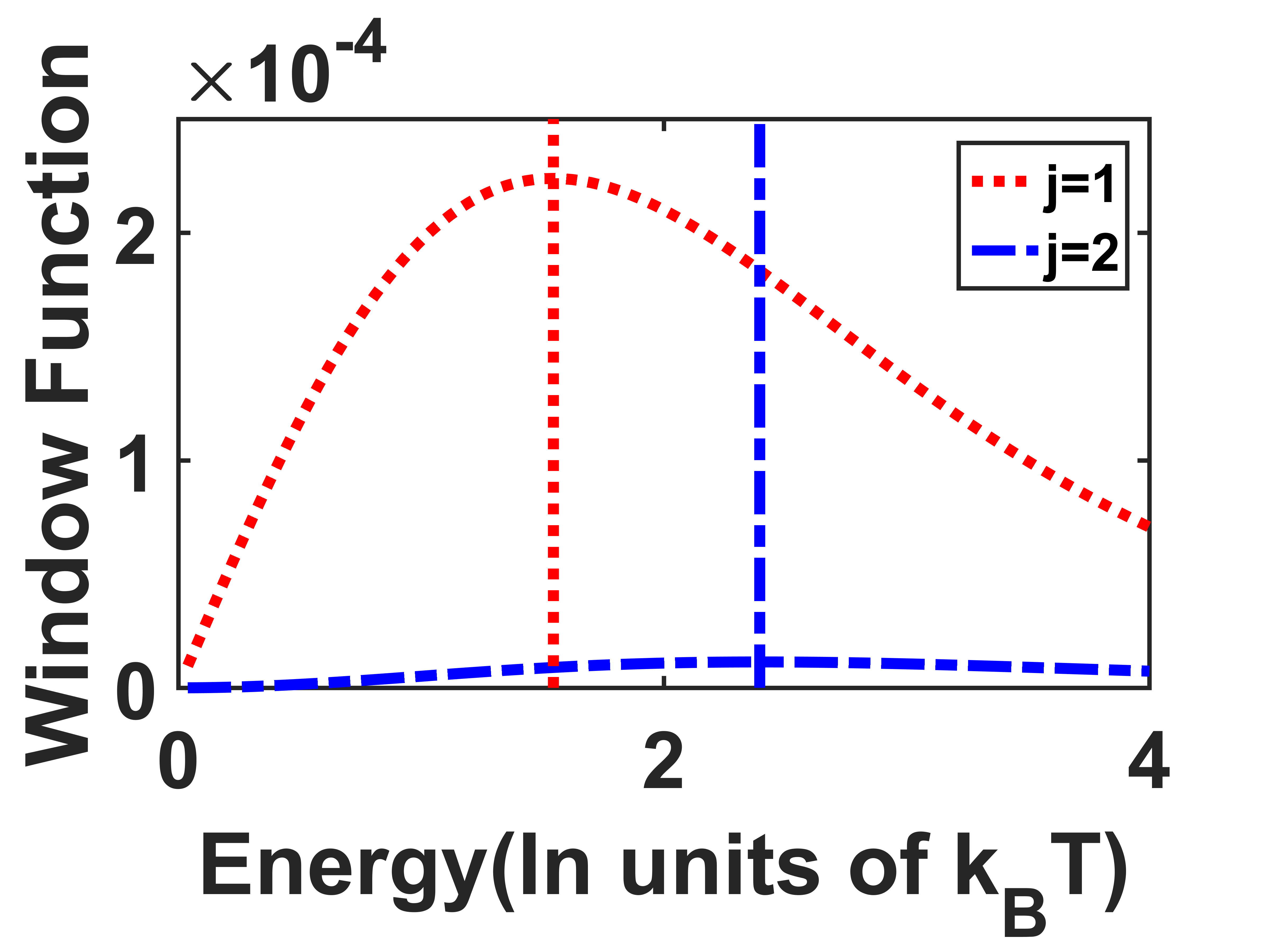}} 
 	\caption{Characteristics of the transport window function. (a) Variation of window function for $j=0$(solid), $j=1$(dot) and $j=2$(dot-dash) as a function of $\eta_F = \frac{E-E_f}{k_BT}$. The extrema (denoted by vertical lines) of the function are further away from $E_F$ as the index $j$ increases. (b) Magnified graph for $j=1$ (dot) and $j=2$ (dot-dash) highlighting the order of magnitude difference between the different window functions for different values of $j$.}
 	\label{fig:window}
 \end{figure}
\indent In a seminal work \cite{sofo}, Mahan and Sofo pointed out that an infinite value of $z_{el}T$ may be asymptotically achieved in the limiting case when the un-broadened DOS and hence the transport distribution \cite{bm} or equivalently, the transmission function tends to a delta distribution. This is typically achieved in a quantum dot system in the limit of vanishing coupling to the contacts. Apart from the thermodynamic interpretation of achieving the Carnot efficiency \cite{linke,esposito1,esposito2,esposito3,bm}, the infinite $z_{el}T$ is attributed to the vanishingly small electronic thermal conductivity \cite{sofo,bm}. This is simply because the delta transport distribution produces a zero variance in energy, leading to a zero electronic thermal conductivity. Thus a resonant enhancement in the electronic DOS is often a sought after route that combines electron filtering with electronic thermal conductivity engineering \cite{heremans,dresselhaus1,dresselhaus2,sofo}, and this will be the primary focus of the current paper.\\
 \indent There has hence been significant interest in molecular \cite{reddy}, quantum dot \cite{Jordan,bm}, super-lattice thermoelectric generators \cite{jordan1,jordan2,akshay}, and also other systems which feature a resonant distortion in the DOS \cite{heremans}, all of which aim to emulate a delta like transmission peak via sharp resonant levels. While the engineering of lattice thermal conductivity concerns the design of interfaces to increase phonon scattering, the aforementioned ideas lay the basic foundations to work with the electronic thermal conductivity by tailoring the electronic DOS. \\
\indent Quantum broadening of energy levels is, however, an inevitable by-product of electronic transport \cite{datta2,bm}, which arises due to coupling with the contacts or electrodes. As a result of broadening, $z_{el}T$ deteriorates drastically as the broadening becomes significant. A schematic depicting this aspect is shown in Fig.~\ref{fig:RTD}(a), and (b), where the broadening of resonant levels in a quantum well is schematically sketched. It must be noted that, traditionally, perfect electron filtering involves a step like transmission function as depicted in Fig.~\ref{fig:RTD}(b), where current flow along only one direction occurs when the Fermi level lies below the band edge. It is hence critical that the broadening function be manipulated, so as to engineer a favorable trade-off between electron filtering and thermal conductivity decrease, should we decide to think along the direction that was proposed in Ref. [5]. In this paper, we propose one such method to tailor the broadening function so as to engineer a decrease in the electronic thermal conductivity and substantiate it with resonant tunnelling devices using graphene nano-ribbons.\\
\indent The schematic of a resonant tunnelling device is shown in Fig.~\ref{fig:RTD}, where the absence of states in the channel at energies below the Fermi level in the transport window leads to no net flow of electrons between cold and hot contacts. Furthermore, a characteristic of such a double barrier structure is an asymmetric broadening of the transmission peaks. This is caused due to an inherent asymmetry between the low lying and higher energy states resulting from the band edge in the contact region \cite{Overhauser}. Our first task is to show that asymmetrically broadened peaks may result in a $z_{el}T$ of 1000 as compared to the step transmission function in typical quantum well structures.\\
\section{Formulation}
\indent In order to formalize the concepts stated above, we employ the transmission formalism in the  linear response regime ~\cite{datta2} to evaluate the transport coefficients from the quantum mechanical transmission function: 
\begin{equation} \label{eq:G}
	\sigma=\frac{2q^2}{h}I_0 ~\Omega^{-1}
\end{equation}
\begin{equation} \label{eq:S}
	S=\frac{k_B}{-q}\frac{I_1}{I_0} ~V/K
\end{equation}
\begin{equation} \label{eq:Ke}
	\kappa_{el}=\frac{T_L2k_B^2}{h}\left(I_2-\frac{I_1^2}{I_0}\right)~W/K,
\end{equation}
where
\begin{equation} \label{eq:Ij}
	I_j=\int \limits_{-\infty}^{\infty} \left(\frac{E-E_F}{k_B T_L}\right)^j \hat{T}(E)\left(-\frac{\partial f_0}{\partial E}\right)dE,
\end{equation}
with $q$ being the electronic charge, $h$ being the Planck's constant, $k_B$ being the Boltzmann constant with $T_L=300K$ being the temperature of the cold contact, $E_F$ being the Fermi level. Here, $\hat{T}(E)$ is the energy resolved transmission function and $f_0$ is the equilibrium Fermi-Dirac distribution function given by $f_0  = \frac{1}{1+e^{\eta_F}}$ where $\eta_F = \frac{E-E_f}{k_BT}$.\\
\indent For multi-moded structures, we calculate the effective transmission with transport along the $\hat{z}$ direction and summing over the transverse modes evaluated by solving the transverse eigenvalue problem \cite{akshay} . For example, in the pure three dimensional (3-D) case \cite{datta1}, this can be written as $\hat{T}(E_x+E_y+E_z)=g_{2D}(E_x+E_y)T(E_z)$, where $g_{2D}(E_{\perp}=E_x+E_y)$ is the two dimensional density of states and is given by $g_{2D}(E_{\perp}) = \frac{m}{\pi \hbar ^2} \theta(E_{\perp}-E_c)$.\\
\section{Results}
\subsection{Toy model: Thermal conductivity engineering}
\begin{figure*}
  	 \subfigure[]{\includegraphics[scale=0.2]{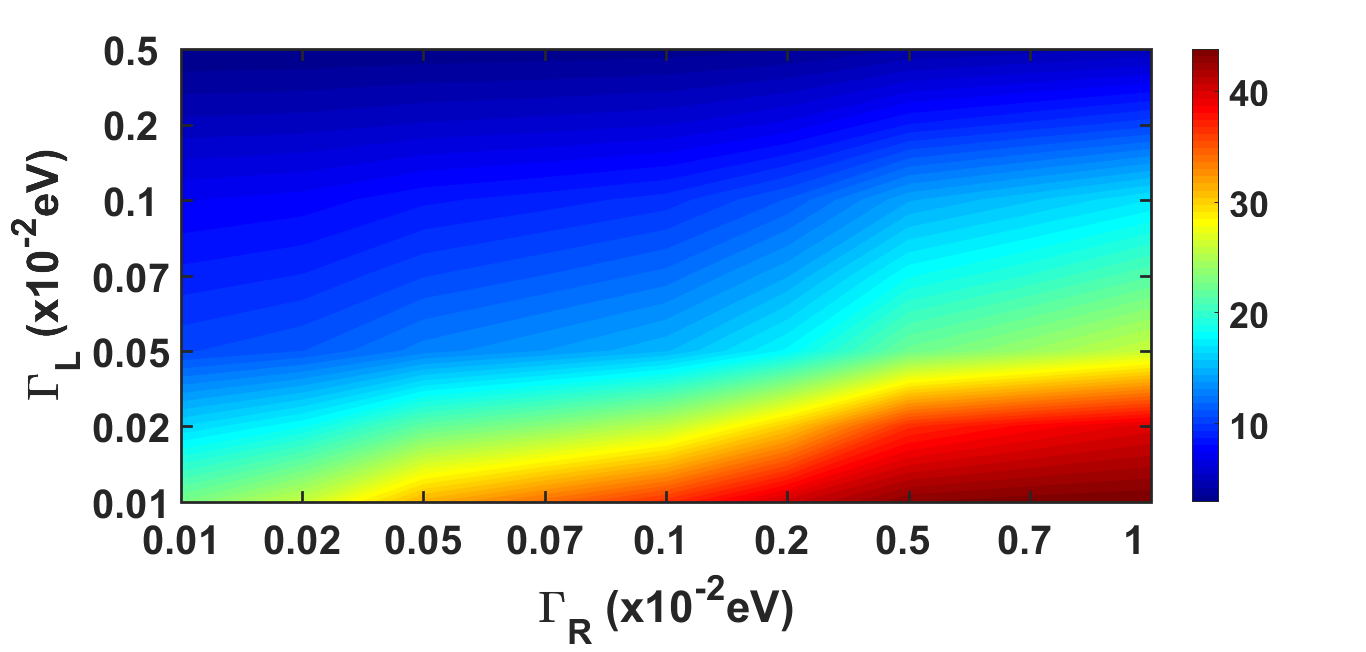}
  	 }
  	 '
  	 \subfigure[]{\includegraphics[scale=0.2]{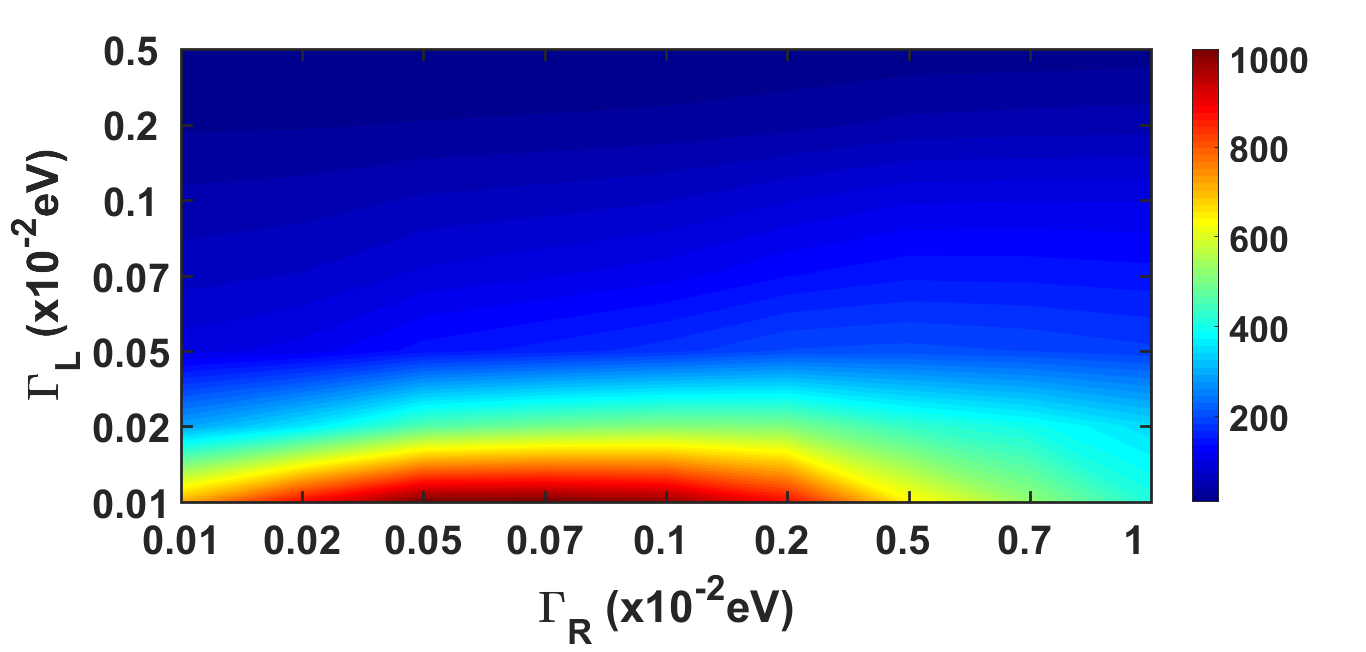}
  	 }
  	  	
  \subfigure[]{\includegraphics[scale=0.2]{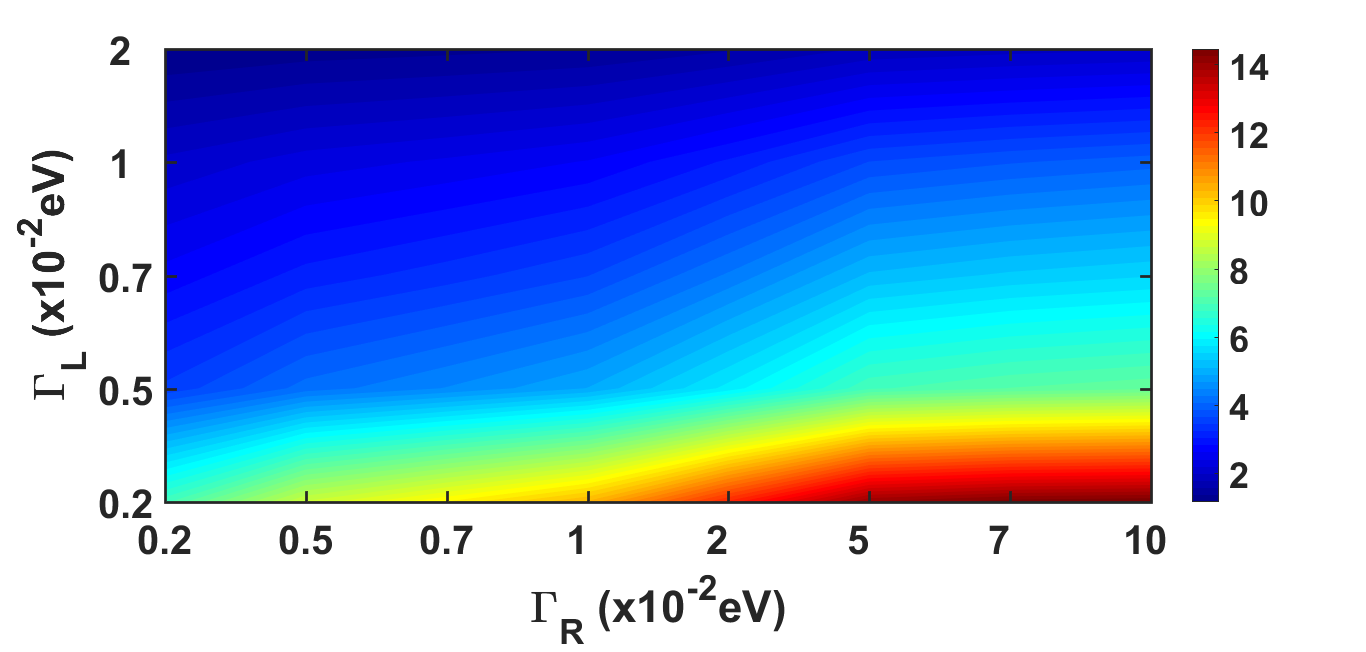}
  }
  '
  \subfigure[]{\includegraphics[scale=0.2]{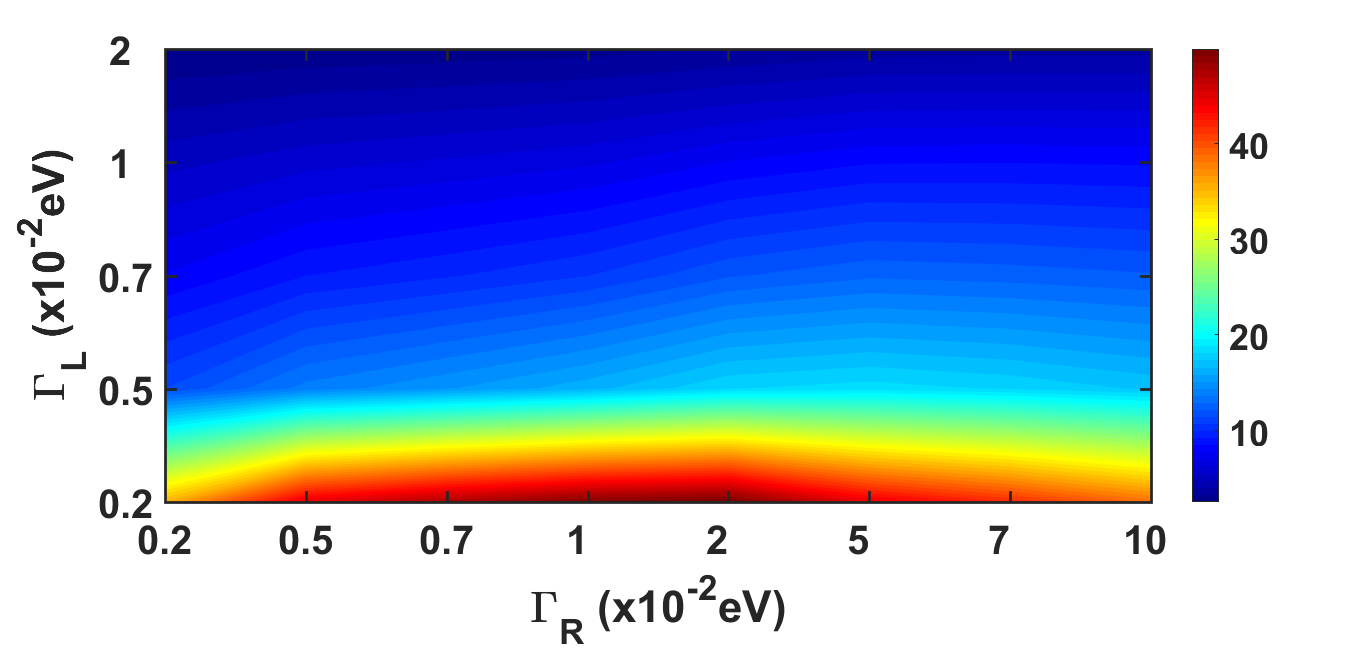}
  }
  \caption{Electronic figure of merit $z_{el}T$ for (a),(c) 3-D structures and (b),(d) 1-D structures, as a function of the left $\Gamma_L$ and right $\Gamma_L$ broadenings in $eV$. (a) and (b) resulting from ultra-low (ideal) ambient broadenings and (c) and (d) resulting from realistic ambient broadenings.}
  \label{fig:TransportModels}
\end{figure*}
The rudiments of engineering the electronic thermal conductivity follow from simple arguments based on the energy distribution of the transport coefficients. The electrical conductivity at a given energy is directly dependent on the difference in the occupation factor of the electrons in the contacts, which at small temperatures maximizes at the Fermi level and dies down sharply upon detuning from it as seen from Fig. \ref{fig:window}(a).  On the other hand, the electronic thermal conductivity, $(\kappa_{el})$, has a strong dependence on how energetically farther away the electronic energy is from the Fermi level $E_F$. Due to the term $\left(\frac{E-E_F}{k_B T_L}\right)^2$  in  \eqref{eq:Ij} being a product of a decreasing function $\left(-\frac{\partial f_0}{\partial E}\right)$ and an increasing function $\left(\frac{E-E_F}{k_B T_L}\right)^2$, its resulting peak is further away from $E_F$ as seen in Fig. \ref{fig:window}(b). Same is the case with the Seebeck coefficient $(S)$. However since the increasing term is linear, the peak in this case is much closer to $E_F$ as seen in Fig. \ref{fig:window}(b). This linear dependence also makes the function odd around $E_F$ thus providing a constraint that the electronic transmission should be on one side of $E_F $. If the transmission is such that only electrons very close to $E_F$ participate in transport, it will result in a drastic decrease in $\kappa_{el}$ with a marginal decrease in the $\sigma$ and $S$.\\
\indent Let us consider an asymmetric transmission function based on the Lorentzian density of states \cite{datta1} given by:
		\begin{equation}\label{eq:Transmission}
		\hat{T}(E_z)= 
		\begin{cases}
		\frac{(\frac{1}{2}\Gamma_L)^2}{(E_z-E_0)^2+(\frac{1}{2}\Gamma_L)^2},& \text{if } E_z\leq E_0\\
		\frac{(\frac{1}{2}\Gamma_R)^2}{(E_z-E_0)^2+(\frac{1}{2}\Gamma_R)^2},              & \text{if } E_z\geq E_0,
		\end{cases}
		\end{equation}
where $\Gamma_{R(L)}$ represent heuristically, the broadening above (below) the central energy $E_0$. Observe that this transmission function for certain values of  $\Gamma_{R(L)}$ only allows transmission of few states near its peak $E_0$ while keeping the transmission function one sided.\\
\indent We plot in Fig.~\ref{fig:TransportModels}(a),(c) and Fig.~\ref{fig:TransportModels}(b),(d), the variation of $z_{el}T$ in three dimensional (3-D) structures and one dimensional (1-D) structures respectively, as a function of $\Gamma_L$ and $\Gamma_R$. For a 3-D structure, we have integrated over the transverse modes, and the 1-D device represents a purely one dimensional transmission. We observe from  Fig.~\ref{fig:TransportModels}(a),(c) and Fig.~\ref{fig:TransportModels}(b),(d), that when $\Gamma_L \neq \Gamma_R$, the electronic figure of merit $z_{el}T$ is larger in comparison to when $\Gamma_L = \Gamma_R$ for both 1-D and 3-D structures. Also, as discussed earlier, an increase in the sharpness of the cut-off at the energy center $E_0$ via a decrease in $\Gamma_L$,  results in an increase in the $z_{el}T$, due to a suppression of reverse electronic flow below the Fermi level. It is also noted by comparing Fig.~\ref{fig:TransportModels}(a),(c) and Fig.~\ref{fig:TransportModels}(b),(d), that a 1-D structure gives rise to a much better performance in $z_{el}T$.
The transmission function proposed here decreases both the electrical conductivity and the electronic thermal conductivity. But due to the special nature of the window functions stated above, the percentage decrease in the three quantities is not the same. The percentage decrease in the Seebeck coefficient is much smaller than the percentage decrease in the electronic thermal conductivity and since we are interested in the ratio of these quantities, this disproportional change in the quantities of the numerator and the denominator of $zT_{el}$ results in its increase. On comparing  Fig.~\ref{fig:TransportModels}(b),(d), we note that while an ultra high $z_{el}T \approx 1000$ may be achieved in 1-D structures with very small ambient broadening, a realistic broadening profile may also result in favorable figures of merit.
We will now take a closer look into the implications of the above on the thermoelectric performance. \\
   \begin{figure}
  	\subfigure[]{\includegraphics[width=1.7in, height=1.2in]{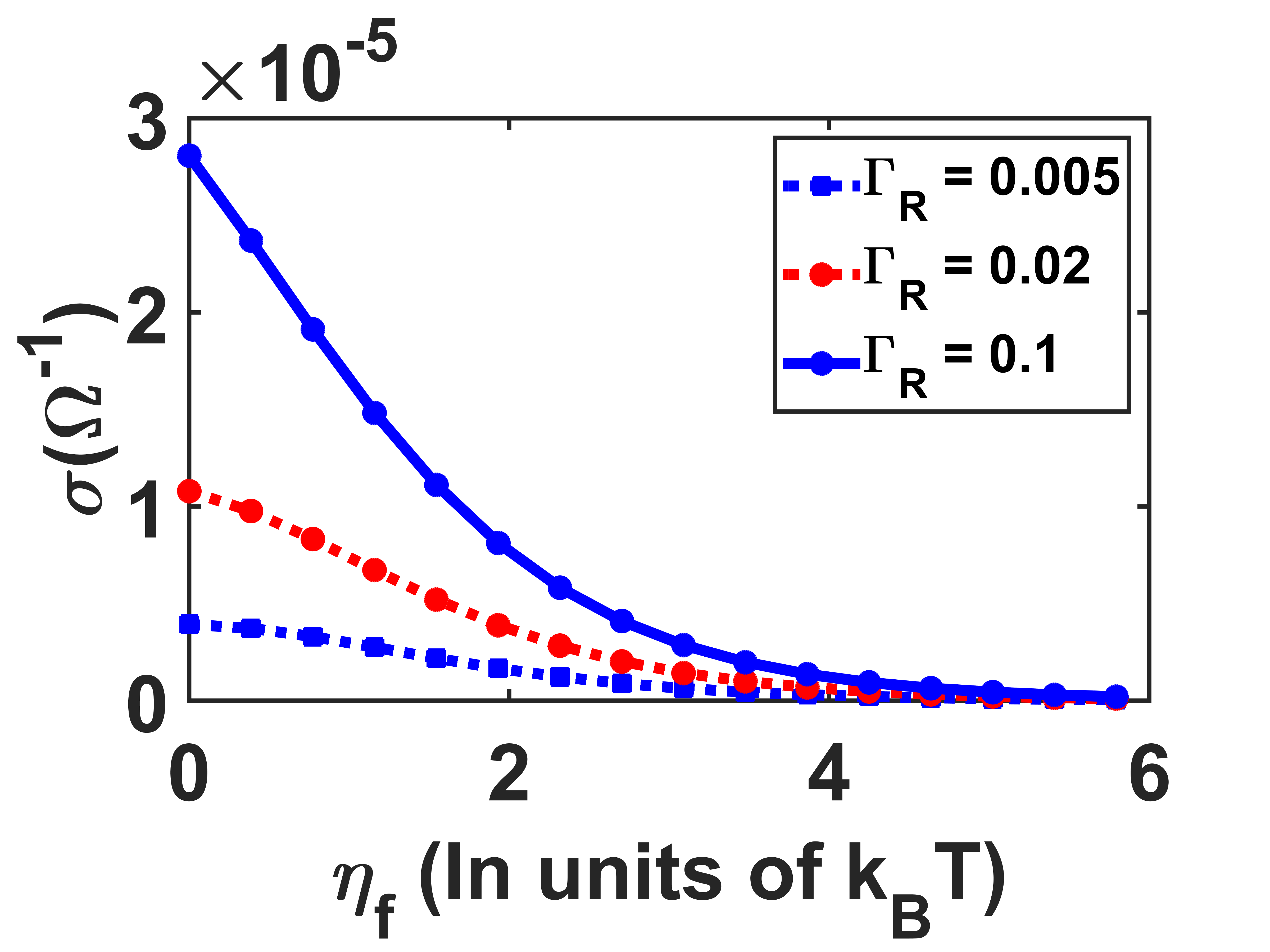}
  	}\subfigure[]{\includegraphics[width=1.7in, height=1.2in]{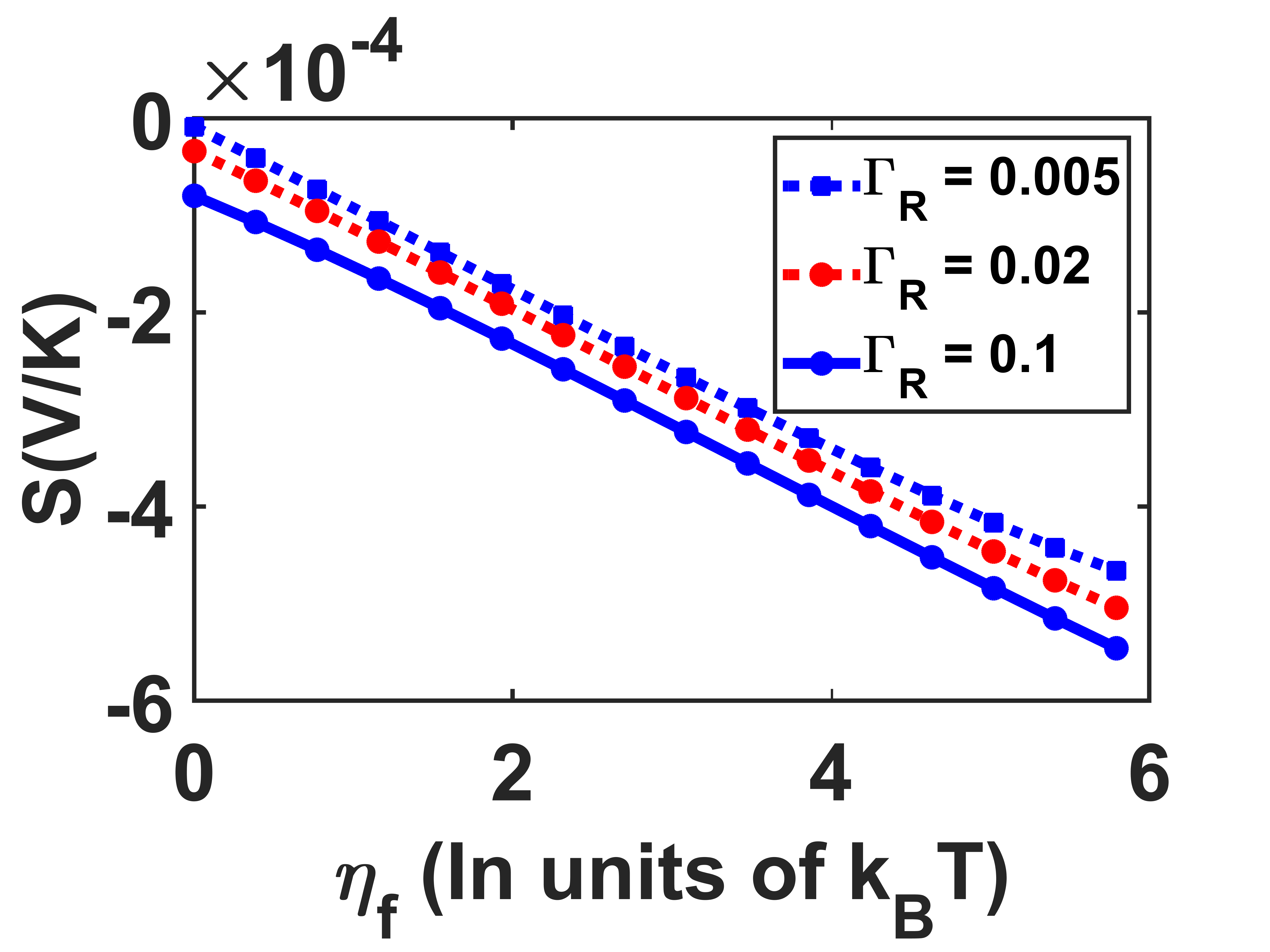}
  }
  \subfigure[]{\includegraphics[width=1.7in, height=1.2in]{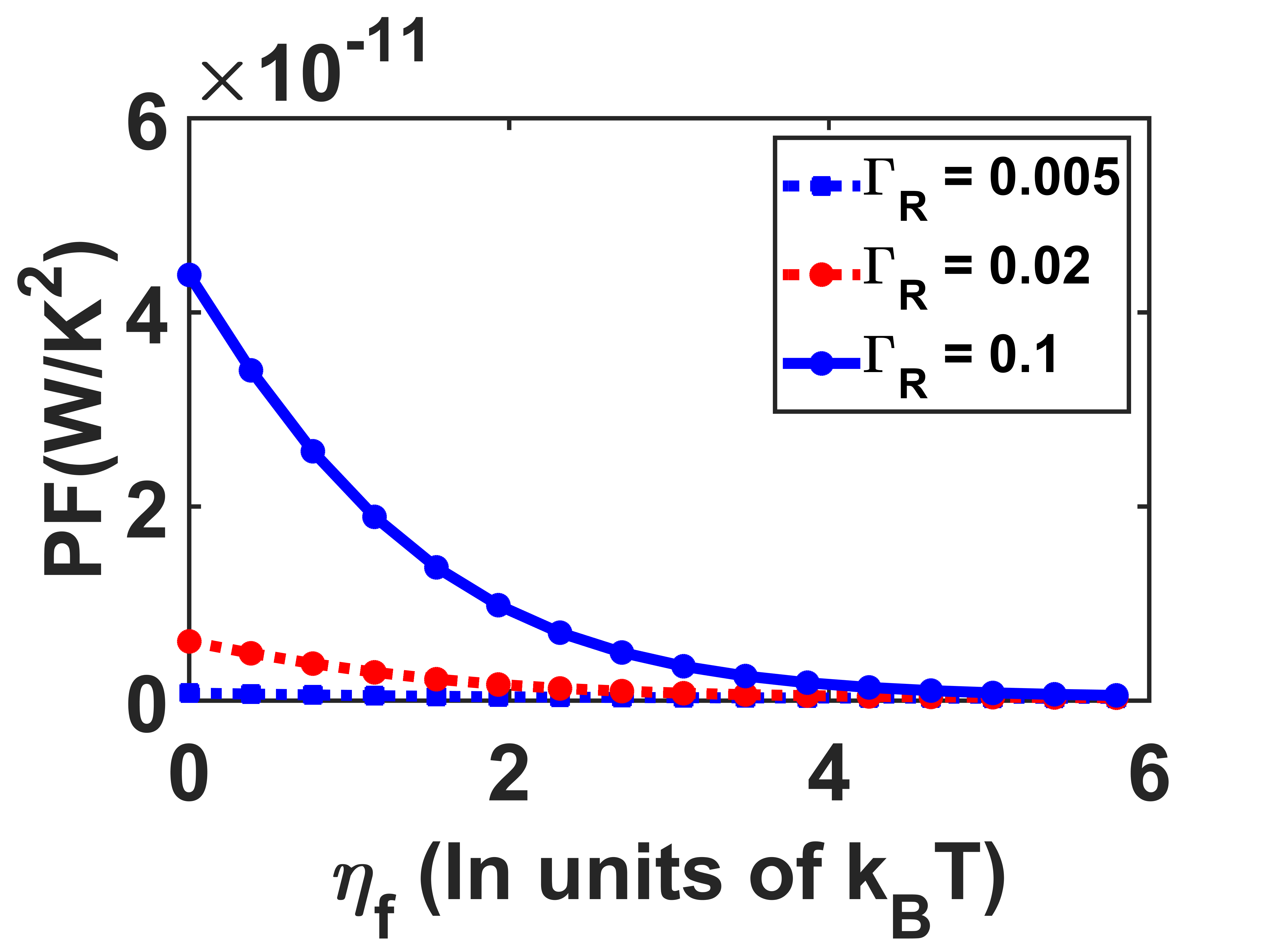}
  }\subfigure[]{\includegraphics[width=1.7in, height=1.2in]{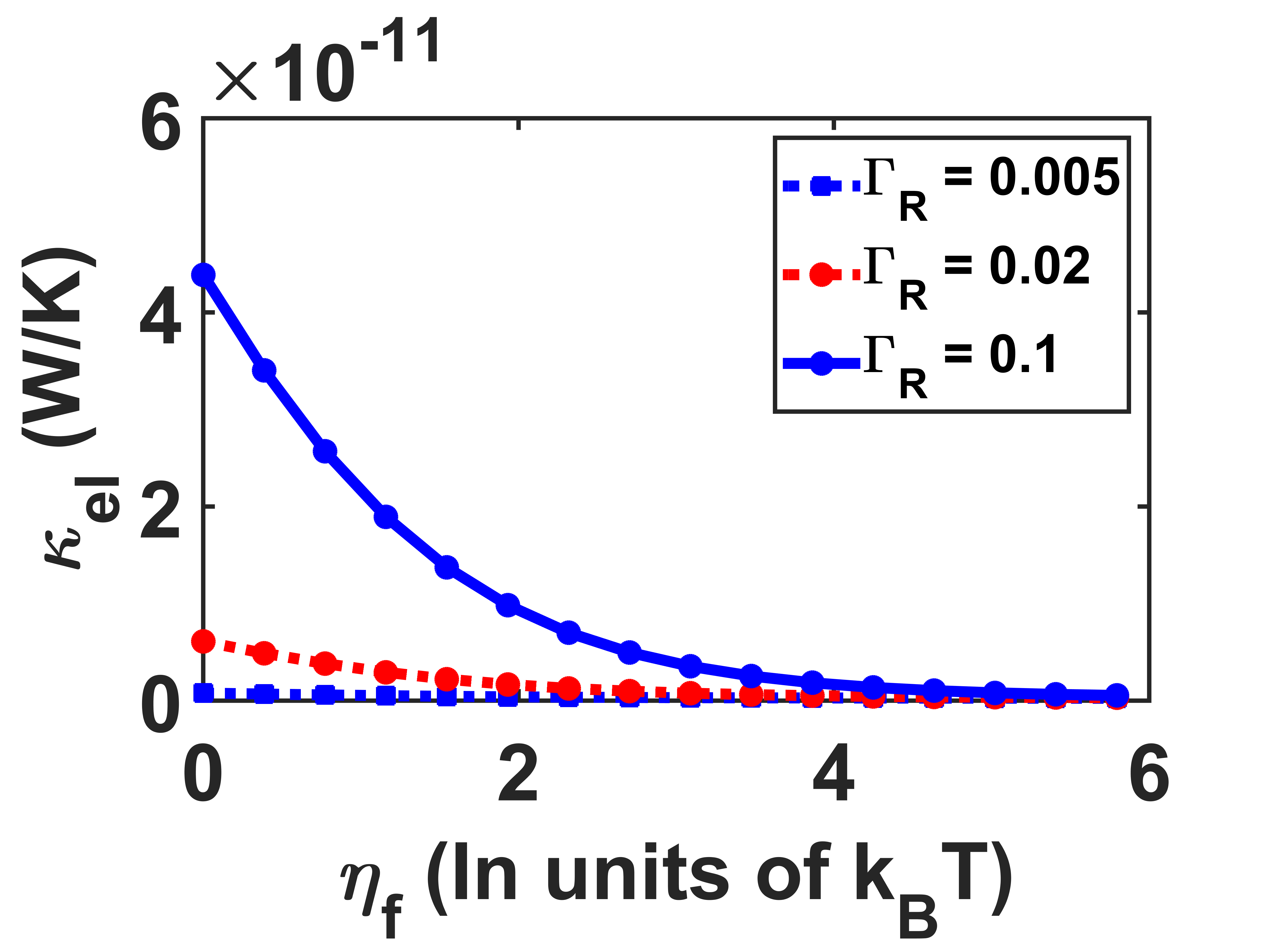}
}

\subfigure[]{\includegraphics[width=1.7in, height=1.2in]{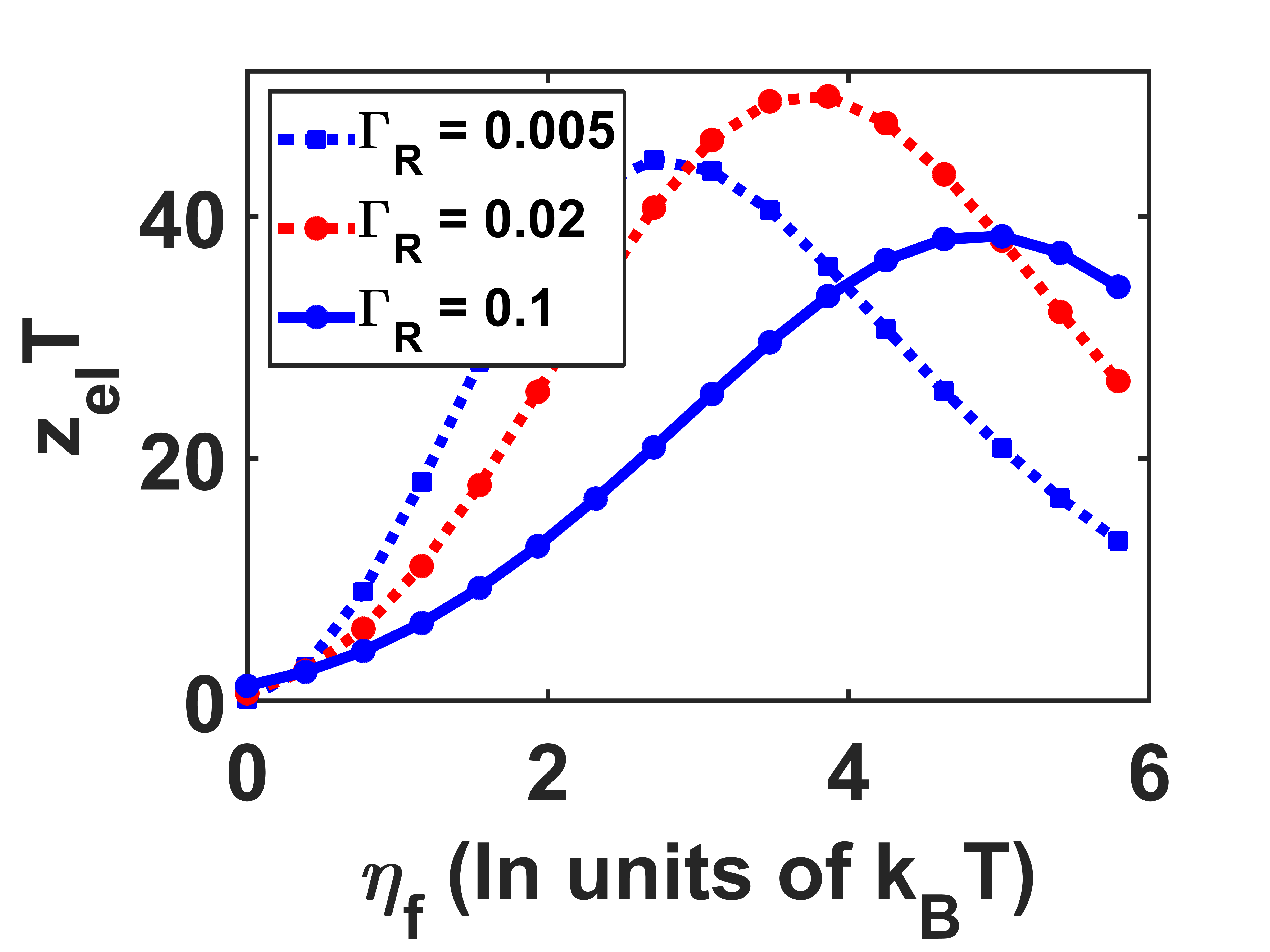}
}
\caption{A closer look at the thermoelectric coefficients for the toy example shown in Fig~\ref{fig:TransportModels}(d). Plots of the variation in (a) Conductivity $\sigma$, (b) Seebeck coefficient $S$ (,c) Power factor $S^2\sigma$, (d) electronic thermal conductivity $\kappa_{el}$, and (e) Electronic figure of merit  $z_{el}T$  as a function of $\eta_f$ with $\Gamma_L = 0.002$ $eV$ for various values of $\Gamma_R$(in units of $eV$).}
\label{fig:idealCoefficients}
\end{figure}
\indent We plot in Fig.~\ref{fig:idealCoefficients}, the thermoelectric transport coefficients as a function of $\eta_F$. We observe that for a fixed $\Gamma_L$ and $\Gamma_R$, the transport coefficients decrease because of a reduced number of conduction channels. For a fixed $\Gamma_L$, upon increasing $\Gamma_R$ , the conductivity $\sigma$ decreases significantly as noted in Fig.~\ref{fig:idealCoefficients}(a). The Seebcek coefficient $S$, however, does not show an appreciable decrease as noted in Fig.~\ref{fig:idealCoefficients}(b), while the electronic thermal conductivity $\kappa_{el}$ decreases as noted in Fig.~\ref{fig:idealCoefficients}(d).  These observations result in the trend of $z_{el}T$ as noted in 
Fig.~\ref{fig:idealCoefficients}(e). We observe that $z_{el}T$ first increases and then decreases with increasing $\Gamma_R$. Any 1D pristine material coupled with contacts will have the transmission to be a broadened step function which is achieved when $\Gamma_R \rightarrow \infty$. So we can conclude from Fig.~\ref{fig:idealCoefficients}(e) that for a given broadening due to coupling from the contacts, the asymmetric broadening results in the maximum $zT_{el}$ that we can have. Therefore, in order to maximize $z_{el}T$, the trade-off between the three transport coefficients noted above explains the peaked behavior of $z_{el}T$.\\
\indent So far, we have demonstrated in Fig.~\ref{fig:idealCoefficients}(e), that it is possible to get an improved $z_{el}T$ via an electronic thermal conductivity decrease. We now need to be able to demonstrate such an effect in realistic super lattice structures. Given that a 1-D structure performs better, we focus on nano-ribbon based structures. In order to design such a structure we need to have control over the band gaps of the constituent materials. One method to control the material properties is to cut two dimensional sheets into ribbons or nano-pattern them \cite{patterning,ga2o3_ribbon,silicene_ribbon2,silicene_ribbon1,atlas,beyond_graphene}. Material properties of such structures are then strongly dependent on the geometry and can be used to form various super-lattice structures \cite{TB,mos2_patterning,QD_graphene}. One such candidate for super lattice structures is graphene nano-ribbons, since the band gaps are dependent on the number of atoms along the width of the ribbon. \\
\subsection{Graphene nano-ribbons}
\indent The band gaps for graphene nano-ribbon (GNR) follow three distinct trends depending on the number of atoms along the width $W=3p , 3p+1 , 3p+2 $, where, $p$ is some integer \cite{TB}. Due to computational complexity in implementing density functional calculations on such super-lattice structures, we have implemented a tight binding Hamiltonian for armchair graphene nano-ribbons using the third nearest neighbor with edge corrections (3NN-EC) described in \cite{TB}. The hopping parameters $t_1=-3.2$ ,$ t_3=-0.3$ , $\delta t_1 =-0.2 $ are used. We ignore the second nearest neighbor hopping as it is shown in \cite{TB} the inclusion of second nearest neighbor interaction only shifts the complete band structure. The transmission function at each energy is then calculated using the ballistic non-equilibrium Green's function (NEGF) formalism \cite{datta1} within the tight binding Hamiltonian framework described above.\\
 \begin{figure}
 	\subfigure[]{\includegraphics[width=3.4in, height=1.2in]{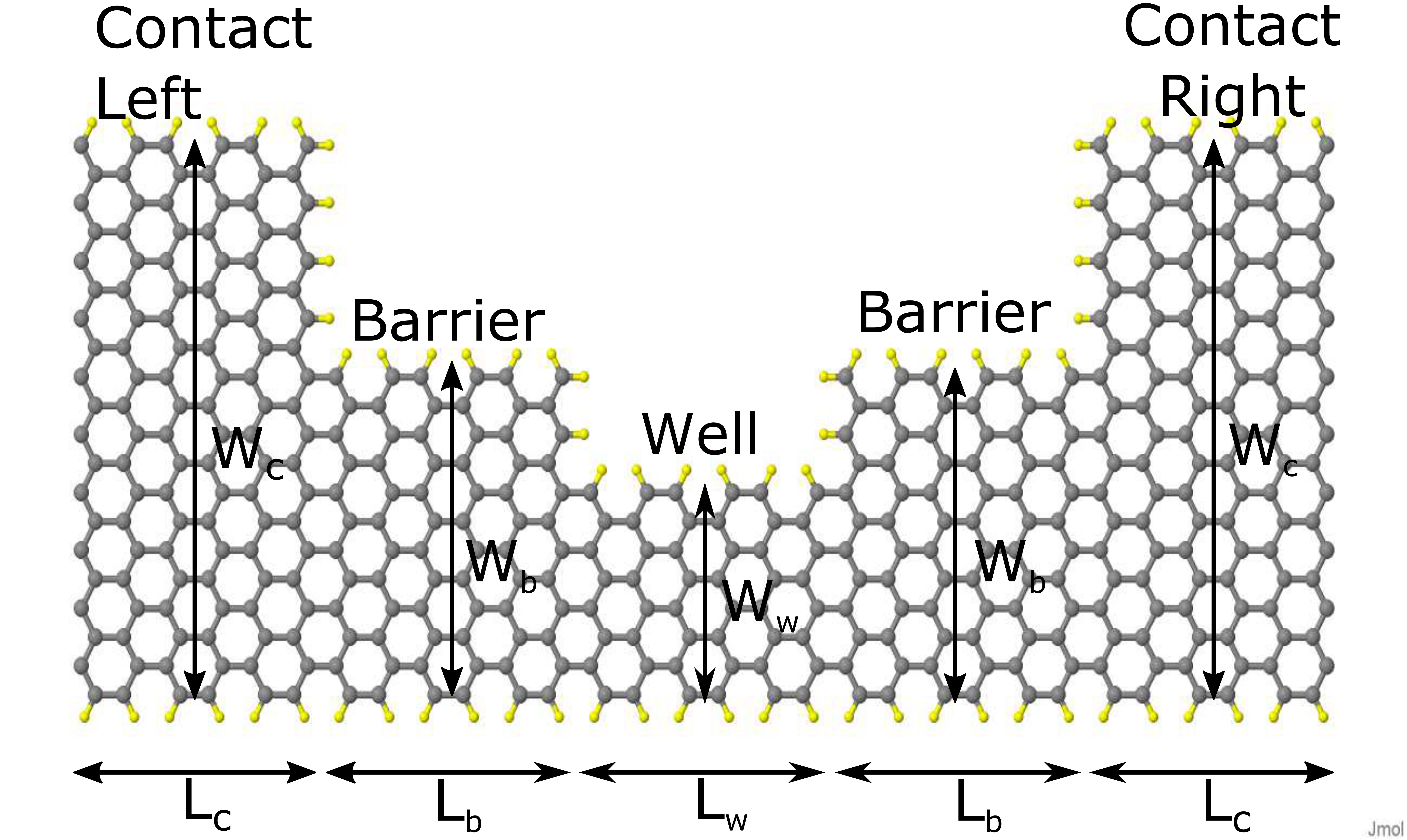}
 	}
 	
 	\subfigure[]{\includegraphics[width=3.4in, height=1.2in]{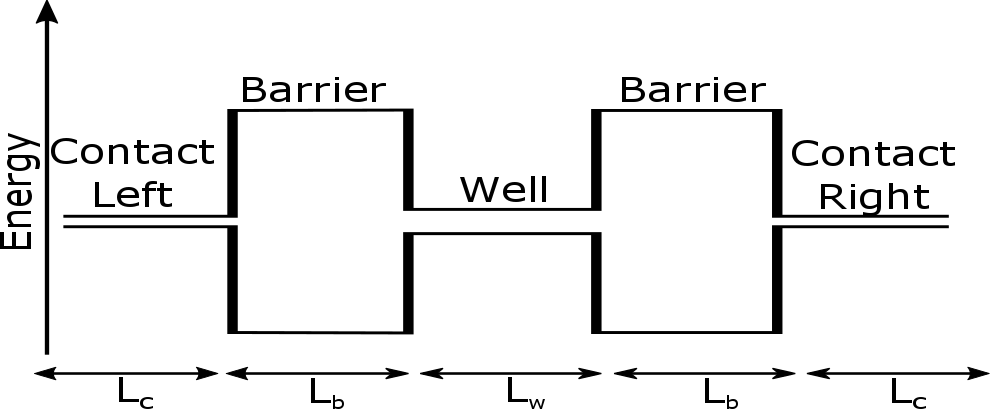}
 	}
 	
 	\caption{Graphene nanoribbon resonant tunnelling device schematics. (a) Schematic with edge dangling bonds passivated with Hydrogen. (b) The band diagram schematic for such a structure. 
 	}
 	\label{fig:RTD_schematic}
 \end{figure}
\indent To validate our TB+NEGF implementation, we have compared our transmission plots with those obtained using density functional theory (DFT) for a set of armchair-GNRs. The results were within acceptable limits. The DFT calculations were done using the Atomistix package\cite{QW1,QW2,QW3}, which is based on the linear combination of atomic orbitals (LCAO) that use the spin polarized Peter-Wang functional within the local density approximation ~\cite{PW-LDA} for the exchange correlation functional and the double-zeta double-polarized basis . The energy grid cut-off for the basis was set to $400~Ry$ with $k$-point sampling of 100 points in the transport direction. Structural relaxation was done to a force tolerance of $16~pN$. The electron temperature was set to $300~K$.\\
\begin{figure}[t]
	\subfigure[]{\includegraphics[width=1.7in, height=1.2in]{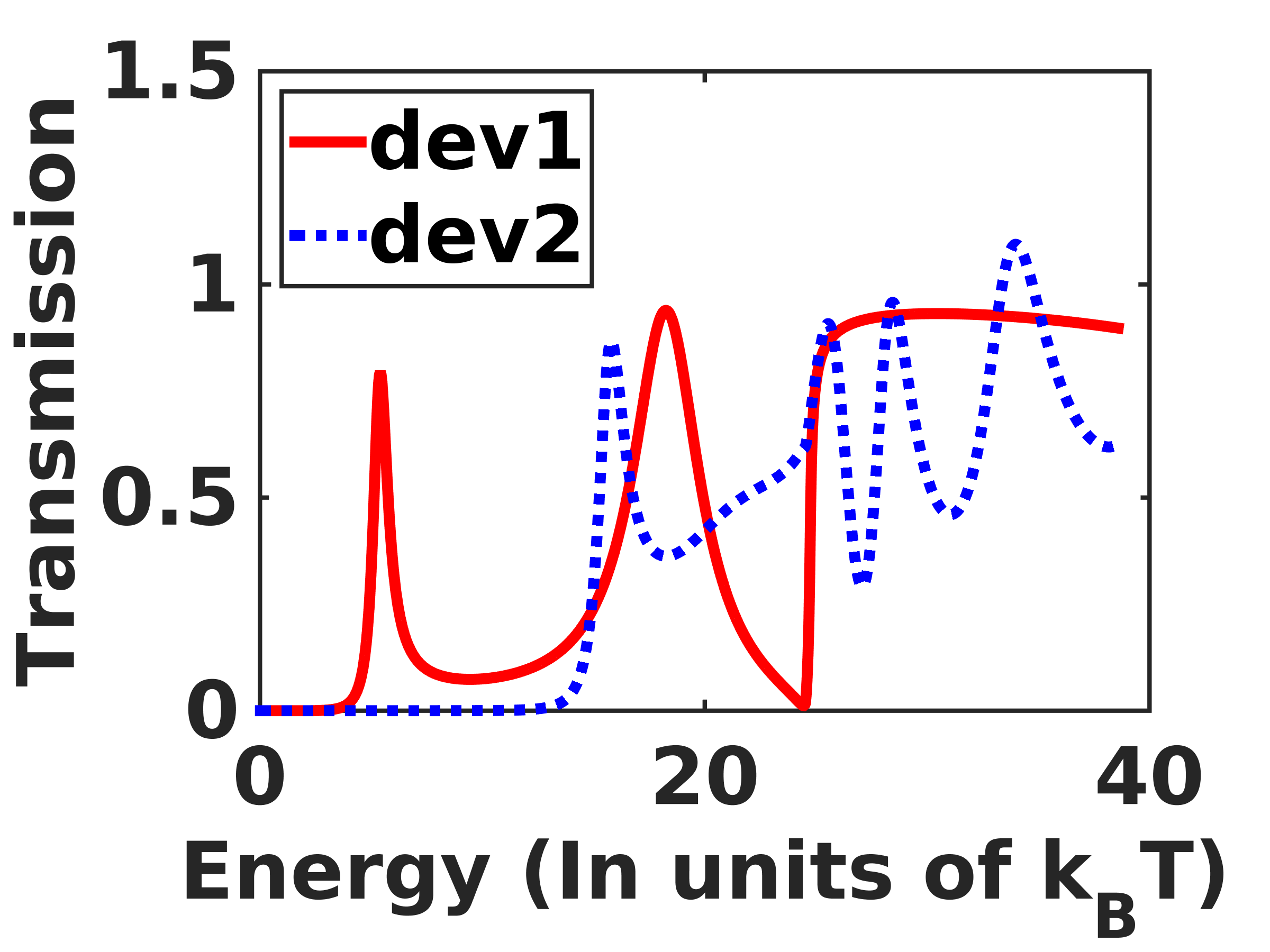}
	}\subfigure[]{\includegraphics[width=1.7in, height=1.2in]{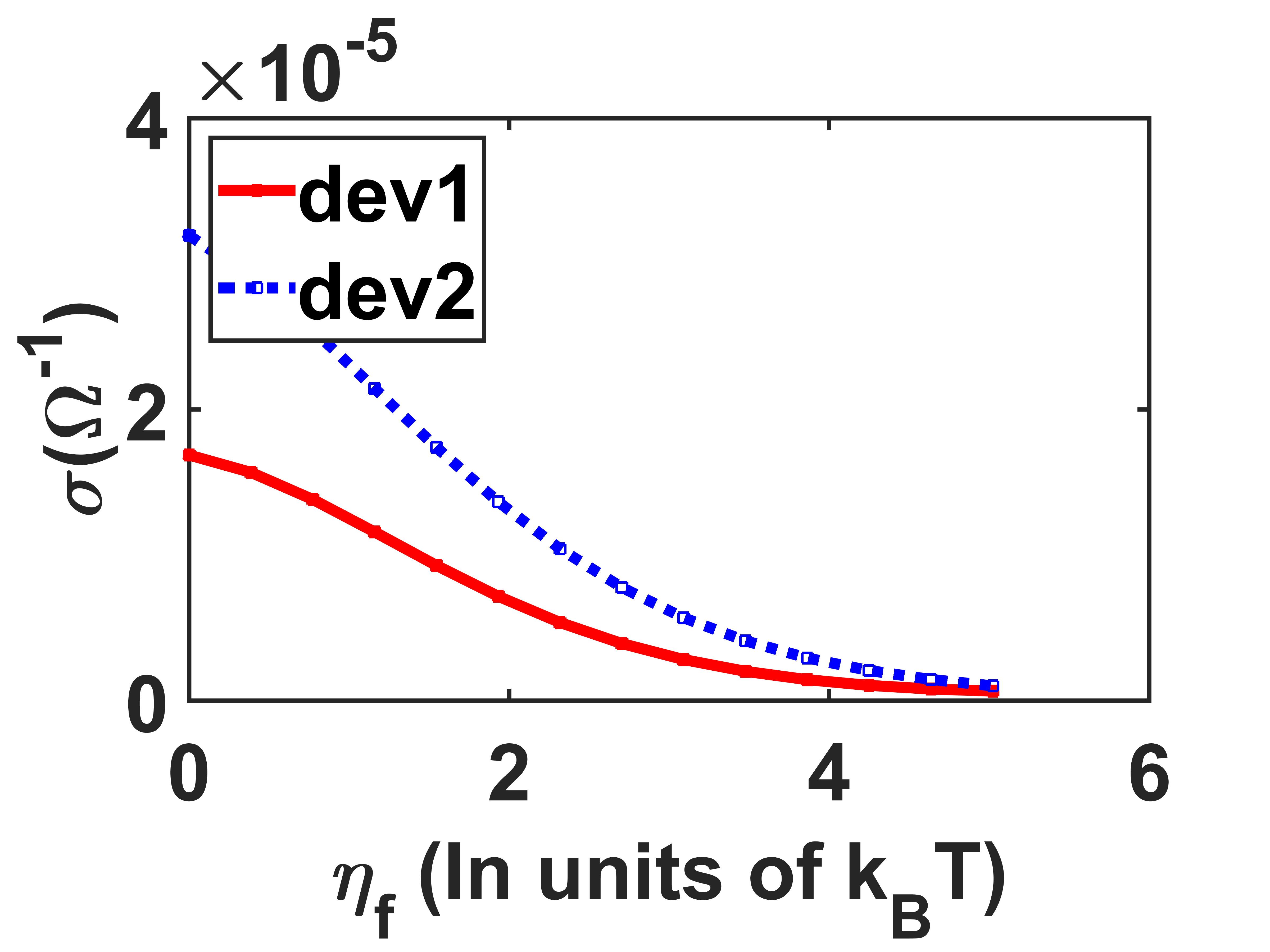}
}
\subfigure[]{\includegraphics[width=1.7in, height=1.2in]{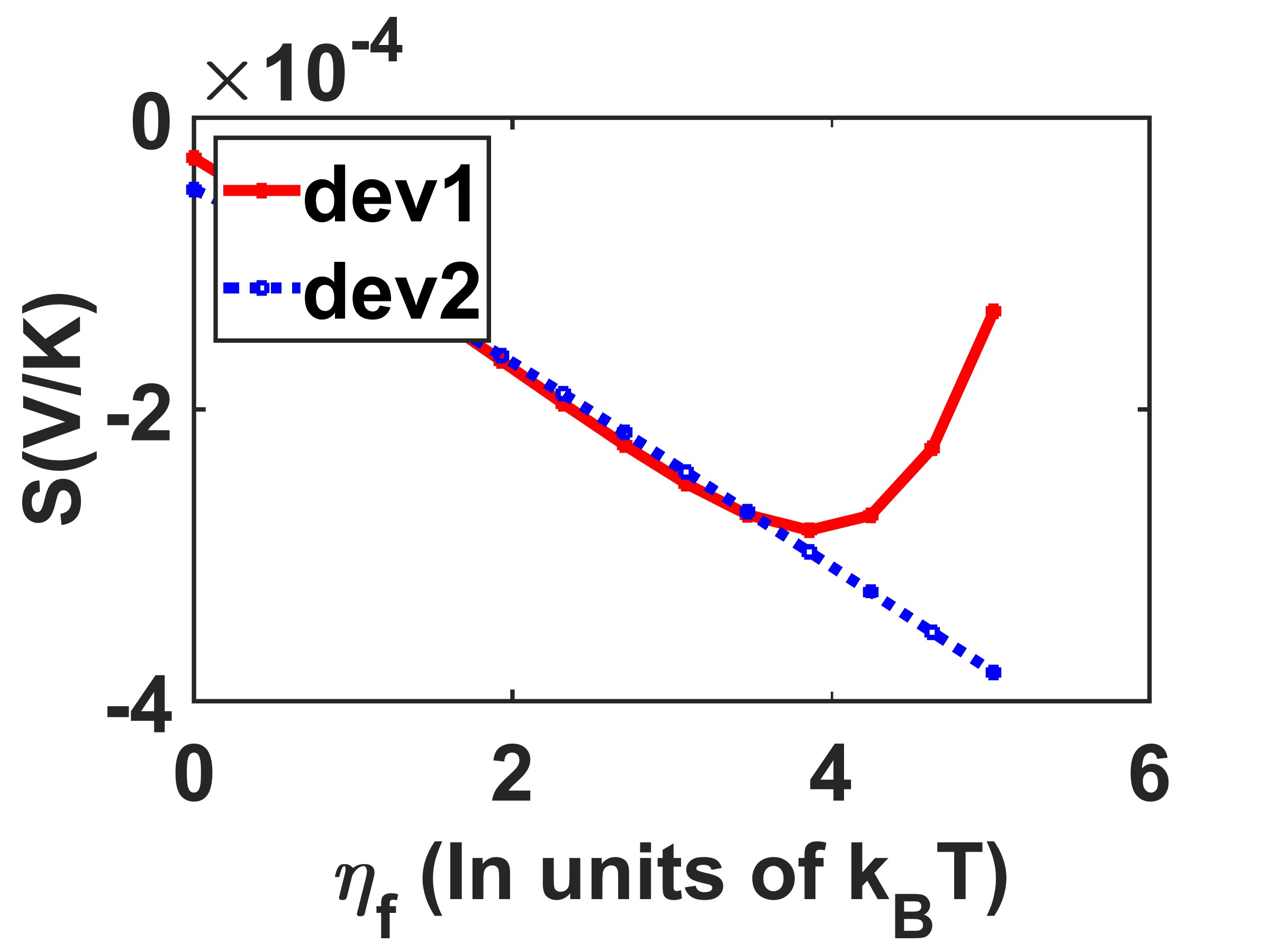}
}\subfigure[]{\includegraphics[width=1.7in, height=1.2in]{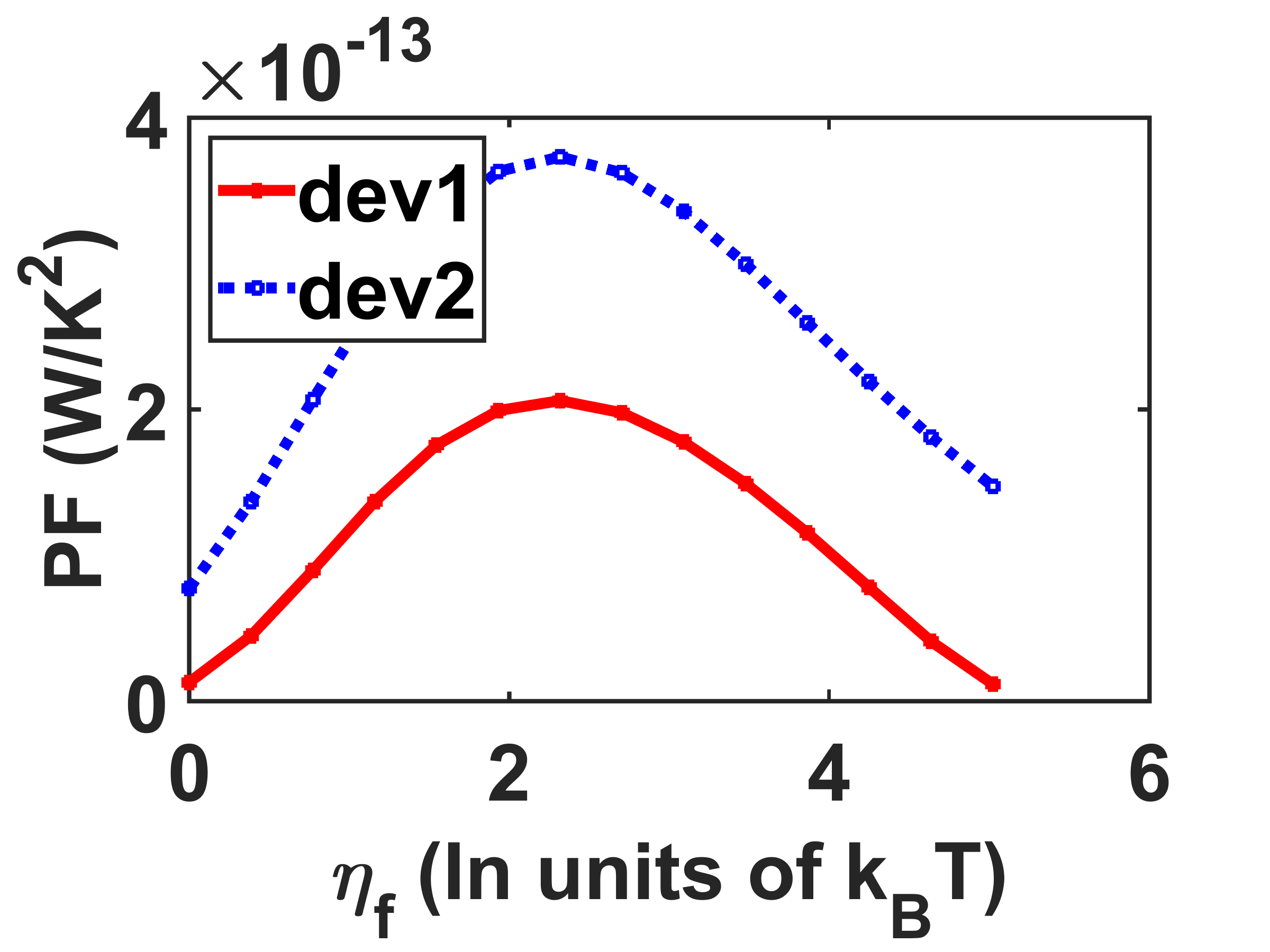}
}
\subfigure[]{\includegraphics[width=1.7in, height=1.2in]{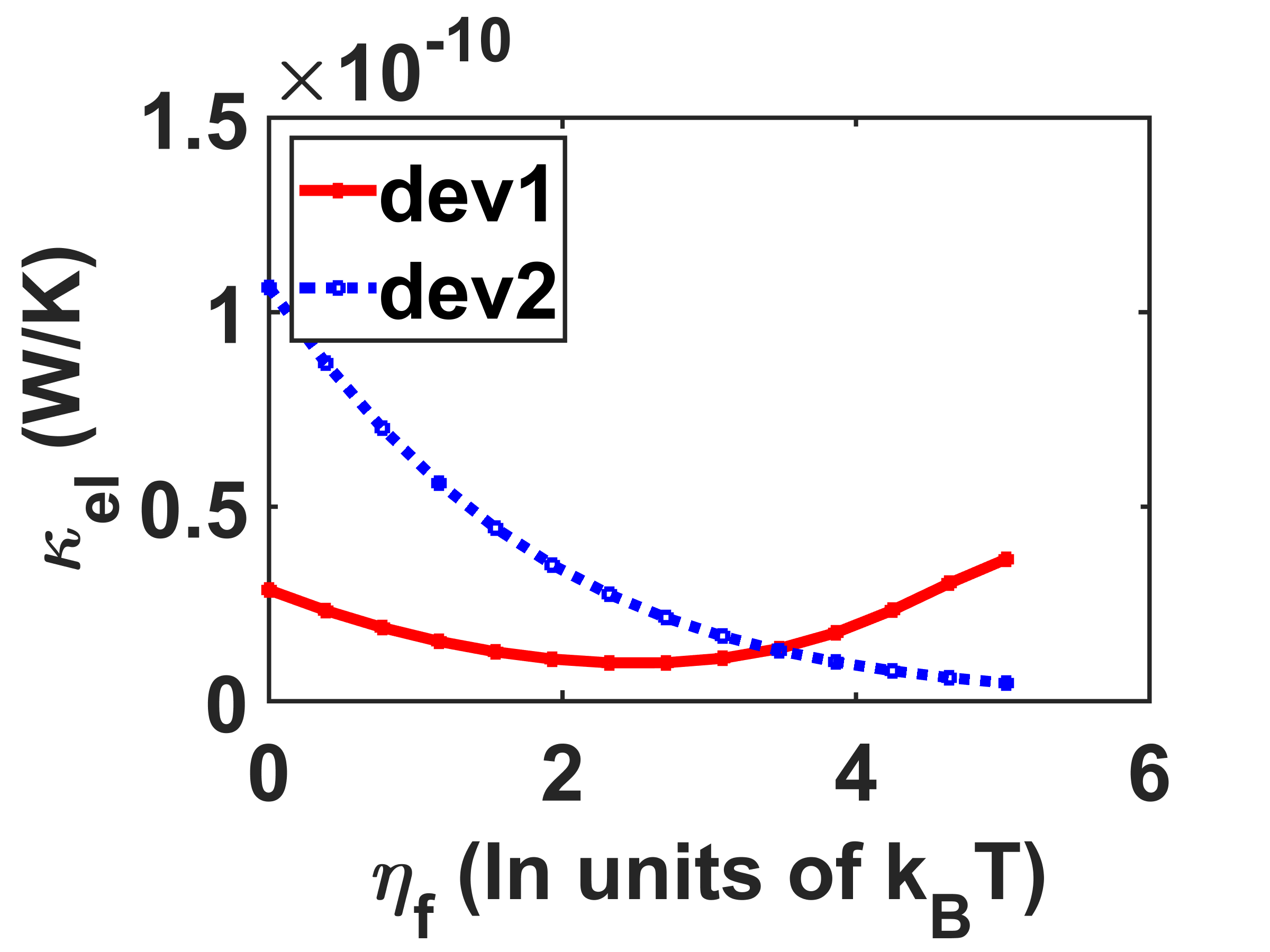}
}\subfigure[]{\includegraphics[width=1.7in, height=1.2in]{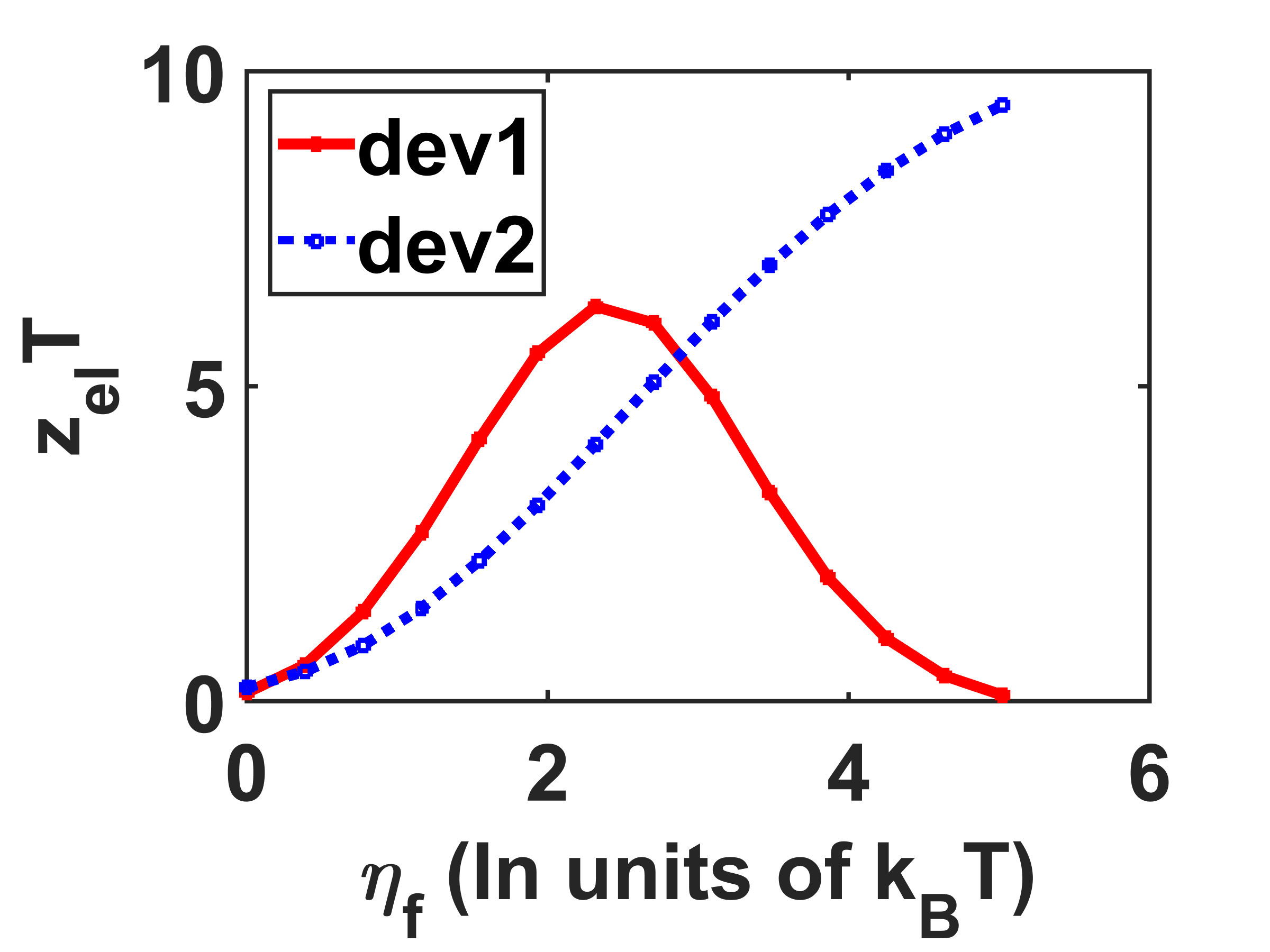}
}
\caption{Thermoelectric performance of realistic GNR resonant tunnelling devices. (a) Transmission plots.  (b) Conductivity $\sigma$, (c) Seebeck coefficient $S$, (d) power factor $S^2\sigma$, (e) lattice thermal conductivity $\kappa_{el}$, and (f) Electronic figure of merit  $z_{el}T$ as function of $\eta_f$. The first device (dev 1) has a geometry $W_{b} = 12$, $W_{w} = 8$, $W_{c} = 20$, $L_{b} = 6$, $L_{w}=6$ and $L_{c}=6$, where as the second device design (dev 2) has a geometry $W_{b} = 10$, $W_{w} = 16$, $W_{c} = 20$, $L_{b} = 6$, $L_{w}=6$ and $L_{c}=6$}
\label{fig:dev_final}
\end{figure}
The transmission function is calculated from the bandstructure obtained using the method described in \cite{datta3}. The transmission spectrum $\hat{T}(E)$ is given by $\hat{T}(E)=T(E)M(E)$. Where transmission $T(E)$ is assumed to be 1 i.e., we assume ballistic transport and $M(E)$ is the Density of modes calculated using 
\begin{equation}
M(E)=\sum \limits _{k_{\perp}} \Theta{\left(E-E_{K{\perp}}\right)},
\end{equation}
where $\Theta{\left(E\right)}$ is the Heaviside step function. The above equation can simply be interpreted as counting the number of bands that cross a given energy in the direction perpendicular to the transport direction. Intuitively, it can be interpreted as counting the number of available parallel paths for electron transfer at a given energy.\\
\subsection{Nano-ribbon-resonant tunnelling structure}
\indent The schematic of the resonant tunnelling structure is shown in Fig.~\ref{fig:RTD}(a), and the band profile is schematically sketched in Fig.~\ref{fig:RTD_schematic}(b). Referring to Fig.~\ref{fig:RTD_schematic}(a), we can tune the transmission function by varying the length of the barrier, $L_b$, the length of the well, $L_w$, the width of the barrier nano-ribbon $W_b$,  and the width of well nano-ribbon $W_w$. Here $L_i$ represents the number of atoms along the length and $W_i$ represents the number of atoms along the width. \\
\indent As described earlier, the asymmetric delta peak can be obtained when the tail of delta distribution merges with the continuum. For this to happen, we should have the energy level which is very close to the top of the well. This can be achieved by having the well very shallow and narrow. As we increase the well depth, the allowed energy levels are deep inside which leads to a decreased broadening of the peak leading to a destruction in the asymmetry. Same is the case when we increase length of the barrier($L_b$).\\
\indent In Fig.~\ref{fig:dev_final}(a), we depict the asymmetric delta function that results from armchair GNR based resonant structures. In both device-1 (dev1), with structural parameters $ W_{b} = 12$, $W_{w} = 8$, $W_{c} = 20$, $L_{b} = 6$, $L_{w}=6$ and $L_{c}=6$, and device-2  (dev2),  with structural parameters $W_{b} = 10$, $W_{w} = 16$, $W_{c} = 20$, $L_{b} = 6$, $L_{w}=6$ and $L_{c}=6$, we observe that the tail of the broadened delta function merges with the higher lying energy levels creating the asymmetry. In Fig.~\ref{fig:dev_final}(c), we see that the variation in Seebeck coefficient $S$ is negligible with change in the device geometry. Similar to what we observed in the toy model in Fig.~\ref{fig:idealCoefficients}(b), we observe a sharp increase in the Seebeck coefficient in Fig.~\ref{fig:dev_final}(d) after $\eta_f = 4 k_BT$. This is because of the contribution of the valance band in the structure of the first device. Similarly, the trends for electrical conductivity $\sigma$ and electronic thermal conductivity $\kappa_{el}$ closely follow those observed in the toy model. Thus this enhancement in electronic figure of merit $z_{el}T$ can be attributed to the drastic decrease in electronic thermal conductivity $\kappa_{el}$ as compared to just the power factor $PF$, thus reinforcing the role of electronic thermal conductivity engineering. At this point, it is worth mentioning that some recent works on phonon scattering across 2-D interfaces \cite{Gunst2011,Mazzamuto2011,Sevincli2010,Feng2016,Xie2016} have re-inforced a lattice thermal conductivity decrease due to the presence of interfaces typical to the structures explored here. Thus we can conclude further based on our work on lowering the electronic thermal conductivity and the aforesaid works on lattice thermal conductivity that the overall $zT$ including the electron and lattice contributions is bound to increase in comparison to pristine systems.
\section{Conclusion}
To conclude, we have suggested an electronic thermal conductivity engineering route to increase the electronic figure of merit $z_{el}T$ of nano-scale structures and demonstrated that this can be done using graphene nano-ribbons. While an idealized device with very small ambient quantum broadening may exhibit an ultra high $z_{el}T \approx 1000$, the realistic device structures explored here present a promising $z_{el}T$ which may be further enhanced using structures designed using other emerging 2-D materials \cite{beyond_graphene} for which the band gaps can be precisely tuned to obtain the optimal broadening. The calculated figure of merit exceeding $10$ in such realistic structures further reinforces the concept and sets a promising direction to use nano-ribbon structures to engineer a favorable decrease in the electronic thermal conductivity. While high $zT$ systems may not typically translate to high output power at high efficiencies \cite{bm,whitney}, such systems may be packaged to obtain a desired output power at  a given efficiency \cite{bm,whitney,Singha,esposito1,esposito2,esposito3}.\\
{\it{Acknowledgement:}} This work was in part supported by the IIT Bombay SEED grant and the Department of Science and Technology (DST), India, under the Science and Engineering Board grant no. SERB/F/3370/2013-2014.

%
%
\bibliography{apssamp}
\end{document}